\documentclass[]{spie}
\usepackage{amsmath,amsfonts,amssymb}
\usepackage[numbers]{natbib}
\usepackage{graphicx}
\usepackage{setspace}
\usepackage{tocloft}
\usepackage[colorlinks=true, allcolors=blue]{hyperref}
\title{SHARP: A spectrograph proposal to fully exploit ELT capabilities and look beyond JWST}

\author[a]{P. Saracco}
\author[a]{P. Conconi}%%%%%%%%%%%%
\author[b]{C. Arcidiacono}%%%%%%%%%%%%%%%%%%%
\author[a,d]{H. Mahmoodzadeh}%%%%%%%%%%%%%%%%%%%
\author[i]{V. Cianniello}%%%%%%%%%%%%%%
\author[g]{A. Gargiulo}%%%%%%%%%%%%%%%%
\author[g]{P. Franzetti}
\author[a]{I. Arosio}%%%%%%%%%%%%%%%%%
\author[a]{L. Barbalini}%%%%%%%%%%%%%%%%%%%
\author[a]{G. Lops}%%%%%%%%%%%%%%%%%%%%
\author[a]{E. Molinari}%%%%%%%%%%%%%%%%%%%
\author[i]{E. Cascone}%%%%%%%%%%%%%
\author[n]{F. D'Alessio}
\author[i]{V. De Caprio}%%%%%%%%%%%%
\author[c]{I. Di Antonio}%%%%%%%%%%%%%%%%%
\author[c]{B. Di Francesco}%%%%%%%%%%%%%%
\author[c]{G. Di Rico}
\author[g]{M. Fumana}
\author[i]{C. Eredia}%%%%%%%%%%%%%%
\author[c]{E. Portaluri}%%%%%%%%%%%%%%%%%%%
\author[n]{F. Vitali}
\author[i]{J. M. Alcal\'a}%%%%%%%%%%%%%%%%%%%
\author[g]{S. Bisogni}%%%%%%%%%%%%%%%%%
\author[l]{R. Bonito}%%%%%%%%%%%%%
\author[m,b,r]{E. Dalla Bont\`a}%%%%%%%%%%%%%%%%%%%%%%%
\author[i]{M. Dall'Ora}%%%%%%%%%%%%%%%%%%
\author[n]{G. Fiorentino}%%%%%%%%%%%%%%
\author[f]{A. R. Gallazzi}%%%%%%%%%%%%%
\author[l]{M. Guarcello}
\author[i]{L. Izzo}
\author[i]{F. La Barbera}%%%%%%%%%%%%%%%
\author[o]{C. Lardo}%%%%%%%%%%%%%%%%%%%
\author[a]{M. Longhetti}%%%%%%%%%%%%%%%%% 
\author[p]{A. Longobardo}%%%%%%%%%%%%%%%%%%%%%
\author[f]{L. Magrini}
\author[g]{C. Mancini}%%%%%%%%%%%%%%%%%%%%%
\author[p]{A. Mura}%%%%%%%%%%%%
\author[n]{E. Piconcelli}%%%%%%%%%%%%%%%
\author[m]{A. Pizzella}
\author[f]{L. Podio}%%%%%%%%%%%%%%%%%%%%%
\author[g]{M. Polletta}
\author[l]{L. Prisinzano}
\author[q,n]{F. Ricci}%%%%%%%%%%%%%%%%%%
\author[i]{V. Ripepi}%%%%%%%%%%%%
\author[f]{V. Roccatagliata}%%%%%%%%%%%%%%%%
\author[h]{E. Vanzella}
\author[g]{G. Vietri}%%%%%%%%%%%%%%%%%%
\author[b]{H. Wang}%%%%%%%%%%%%%%%%%%

\affil[a]{INAF - Osservatorio Astronomico di Brera, Via Brera 28, Milano, Italy, 20121}
\affil[b]{INAF - Osservatorio Astronomico di Padova, Vicolo Osservatorio 5, Padova, Italy, 35122}
\affil[c]{INAF - Osservatorio Astronomico d'Abruzzo,Via Mentore Maggini, Teramo, Italy, 64100}
\affil[d]{Politecnico di Milano, Piazza Leonardo da Vinci 32, Milano, Italy, 20133}
\affil[f]{INAF - Osservatorio Astrofisico di Arcetri, Largo E. Fermi 5, Firenze, Italy, 50125}
\affil[g]{INAF - IASF, Via Alfonso Corti 12, Milano, Italy, 20133}
\affil[h]{INAF - Osservatorio di Astrofisica e Scienza dello Spazio, via Gobetti 93/3, Bologna, Italy, 40129}
\affil[i]{INAF - Osservatorio Astronomico di Capodimonte, Salita Moiariello 16, Napoli, Italy, 80131}
\affil[l]{INAF - Osservatorio Astronomico di Palermo, Piazza del Parlamento 1, Palermo, Italy, 90134}
\affil[m]{Universita' degli Studi di Padova, Via F. Marzolo 8, Padova, Italy, 35131}
\affil[n]{INAF - Osservatorio Astronomico di Roma, Via Frascati 33, Monte Porzio Catone (RM), Italy, 00078}
\affil[o]{Universita' degli Studi di Bologna, via Piero Gobetti 93/2, Bologna, Italy, 40129}
\affil[p]{INAF - IAPS, Via del Fosso del Cavaliere 100, Roma, Italy, 00133}
\affil[q]{Dipartimento di Matematica e Fisica, Universit\'a Roma Tre, Via della Vasca Navale 84, Roma, Italy, 00146}
\affil[r]{Jeremiah Horrocks Institute, University of Central Lancashire, Preston, PR1 2HE, UK}

\authorinfo{Further author information: (Send correspondence to P. Saracco)\\
E-mail: paolo.saracco@inaf.it, SHARP website: http://sharp.brera.inaf.it}

\begin{document}
\maketitle

\begin{abstract}
The Extremely Large Telescopes (ELTs), with their large apertures and cutting-edge Multi-Conjugate Adaptive Optics (MCAO) systems, promise to deliver data sharper and deeper than even the James Webb Space Telescope (JWST) across large fields. SHARP is a concept study for a near-IR (0.95--2.45 $\mu$m) spectrograph specifically designed to fully exploit the collecting area and angular resolution capabilities of the upcoming ESO's ELT. The instrument concept is driven by the goal of tackling the most important questions in astrophysics and cosmology, bridging the gap between the local and distant Universe. This requires versatility to accommodate diverse observational needs. SHARP (http://sharp.brera.inaf.it) is composed of two main units: NEXUS, a Multi-Object Spectrograph (MOS) optimized for detecting the faintest sources, and VESPER, a multi-object Integral Field Unit (multi-IFU) designed for brighter ones. This presentation provides an overview of the scientific design drivers, the solutions developed to meet them, and the resulting optical design that achieves the required performance.
\end{abstract}


\section{Introduction}
\label{sec:intro}  
The upcoming generation of ground-based Extremely Large Telescopes (ELTs) will peer into the Universe with unparalleled clarity, delivering sharper and deeper data than the James Webb Space Telescope (JWST) \cite[JWST, Ref.][]{gardner23}. 
These capabilities are fundamentally due to their large apertures, providing a massive photon collecting area, coupled with advanced multi-conjugate adaptive optics (MCAO) systems designed to uniformly correct large fields for atmospheric turbulence. 
The ELTs will achieve data significantly sharper than JWST, up to four times in the case of the 
Giant Magellan Telescope \cite[GMT, Ref.][]{matt06,burgett24} 
($\sim$25 m diameter), and six times sharper with the ESO's Extremely Large Telescope
\cite[ELT,][]{gilmozzi07,vernet24}
Specifically, ESO's ELT is positioned to set new standards in astronomical observations, 
thanks to its 39m diameter primary mirror and its advanced MCAO system, MORFEO \citep[Ref.][]{ciliegi21,ciliegi24}. 
MORFEO corrects an area with a diameter of 160 arcsec, an AO system that allows supported 
instruments to explore the Universe with unmatched depth and clarity over large scales.

A spectrograph that fully exploits the capabilities of an MCAO system like MORFEO at the ELT, namely, 
the largest AO corrected area and the faintest achievable flux limit, is expected to surpass 
NIRSpec at JWST in terms of spectral and angular resolution. 
SHARP\footnote{http://sharp.brera.inaf.it} is a concept study for a multi-mode near-IR spectrograph specifically designed to leverage these exact capabilities. It would provide an effective spectroscopic follow-up of the deep and 
high-angular-resolution crowded fields expected from first-light imager MICADO \citep[Ref.][]{davies21,davies18}. 
Developed as a candidate for future ESO instrumentation calls, SHARP is engineered to offer an angular resolution of $\sim$30-35 mas and a wide range of spectral resolutions ($R=300, 2000, 6000, 17000$) 
that go well beyond those provided by JWST. 

Throughout this paper, we adopt H$_0$=70 km/s/Mpc, $\Omega_m$=0.3 and $\Omega_\Lambda$=0.7.

\section{Scientific context and drivers}
\label{sec:science}
The development of a spectrograph like SHARP is driven by the need to understand how baryons assembled to form the first stars, galaxies, and cosmic structures, and how these systems subsequently evolved over cosmic time. 

\subsection{Formation and Evolution of Galaxies}
A deep understanding of the galaxy life-cycle requires dissecting star formation processes, 
the mechanisms at play, environmental effects, and the physical processes that quench star formation. 
Deep observations have revealed numerous massive galaxies at high redshift ($z>3$), challenging our understanding of galaxy formation physics. These galaxies appear too massive, too numerous, and too early in cosmic timescales compared to theoretical expectations. 

JWST has detected massive quiescent galaxies up to $z\sim5$, when the Universe was younger than 1.2 Gyr. These include galaxies whose stellar populations are either extremely old, forming just millions of years after the Big Bang, or much younger, forming right before the epoch of observation  \cite[e.g.,][]{glazebrook17,tanaka19,saracco20apj,deugenio20,antwi-danso25,carnall24,glazebrook24}. 
Given their high stellar masses ($\log(M_*/M_\odot) \sim 11$), the required star formation rates 
must exceed hundreds of solar masses per year. 
These extreme values are highly challenging under the conditions of the early Universe due to 
the lower cooling efficiency of pristine hydrogen and the intense UV radiation density. 
Furthermore, the presence of these massive galaxies in a quiescent state implies that an 
extremely efficient and abrupt quenching mechanism must have taken place. 

\subsection{Scaling Relations and the Stellar Initial Mass Function}
Galaxies follow key empirical scaling relations that link their structural and kinematic properties,
such as effective radius $R_e$ and velocity dispersion $\sigma_e$, to the properties of
their stellar populations such as age and metallicity \citep[e.g.,][]{gallazzi05, saglia16, beifiori17,jorgensen13,holden10}.

Up to $z\sim1.5$, more massive galaxies consistently host older and more metal-rich stellar populations, regardless of the environment \citep[e.g.,][]{jorgensen14,saracco23,thomas05,thomas10,choi14,mcdermid15}. This creates a conceptual tension between the extended time required for metal enrichment and the 
rapid star formation implied by these old stellar ages. 
Furthermore, a mass-dependent dispersion in these properties points to a complex evolutionary 
picture \citep{bevacqua24}, demanding highly detailed measurements of galaxy properties to be 
fully understood.  

A central unknown in these studies is the stellar Initial Mass Function (IMF). 
Traditionally assumed to be universal, recent evidence suggests the IMF varies across different environments\citep[e.g.,][]{hopkins18,smith20}. 
Studying IMF gradients within galaxies can help distinguish between \textit{in situ} star 
formation and mass assembly via mergers \citep[e.g.,][]{conroy12,labarbera13,spiniello14,labarbera19,parikh18}. 
To probe IMF radial gradients at high redshift necessitates sampling scales smaller than 500 pc. 
Such studies remain beyond reach with JWST due to its moderate spatial resolution 
(0.1 arcsec pixel size).

\subsection{Population III Stars}
Population III (Pop III) stars are the theoretical first generation of stars, forming from pristine hydrogen and helium with zero metallicity. 
Detecting them is one of the holy grails of modern astrophysics. 
The primary observational marker for these ultra-hot systems is expected to be the HeII $\lambda$1640 emission line (the recombination n=3-->2 line) and no metal lines \citep[e.g.,][]{nakajima22,katz23,vanzella23}. 
Since these stars have extremely short lifetimes, the probability of detecting them is maximized by 
looking at the earliest epochs ($z>8-9$). 
The detection of the HeII line at these redshifts strictly requires high-sensitivity near-infrared spectroscopy ($\lambda > 1.6 \mu$m) and fine angular resolution to maximize the signal-to-noise ratio 
over the compact ($\sim$200 pc) primordial systems (see Saracco et al. 2026, for a detailed derivation 
of the expected line flux).

\subsection{Star Formation in Extreme Environments}
In the local Universe, most of our knowledge of low-mass star formation comes from nearby 
star-forming regions \citep[$d < 500 \text{ pc}$; e.g.,][]{evans09, dunham15}. 
However, to understand how environment affects star and planet formation, we must study 
Young Stellar Objects (YSOs) in distant, extreme environments (e.g., low-metallicity regions in the Magellanic Clouds or outer Milky Way). 
A highly debated topic is whether low-metallicity environments accelerate the dispersal of circumstellar disks \citep[e.g.,][]{yasui21} due to more efficient photo-evaporation driven by massive stars \citep[e.g.,][]{2022EPJP..137.1132W}. 
Resolving these dense, distant star-forming regions to measure the IMF and disk lifetimes requires exceptional angular resolution to disentangle crowded fields and high multiplexing to build statistically significant samples.

\section{Main requirements}
\label{sec:requirements}
Ultimately, all the specific issues summarized above converge toward the more general and
fundamental issue on how baryons assembled to form galaxies, that is what regulates 
the star formation efficiency, what is able to (sometimes abruptly) quench it and what 
keeps quenched a galaxy.
To tackle these diverse astrophysical issues, SHARP was designed around several
observational requirements.
\subsection{Angular Resolution (30-35 mas)}
To investigate the baryons assembly in galaxies, regions comparable to the 
physical sizes of Giant Molecular Clouds (GMCs) should be resolved over the whole cosmic time. 
Indeed, GMCs containing even more than $10^6$ M$\odot$ of molecular gas, are  

- the main star-forming units

- the main metal production units

- galaxy rotation and kinematics tracers

- the cradles of globular clusters (GC) and, possibly,

- cradles of the first PopIII systems.

\begin{figure}[htbp]
    \centering
    \includegraphics[width=8truecm,height=6truecm]{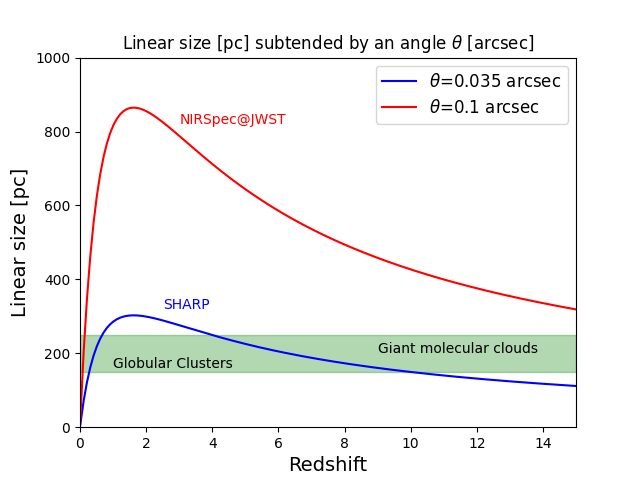}
     \includegraphics[width=9truecm,height=6truecm]{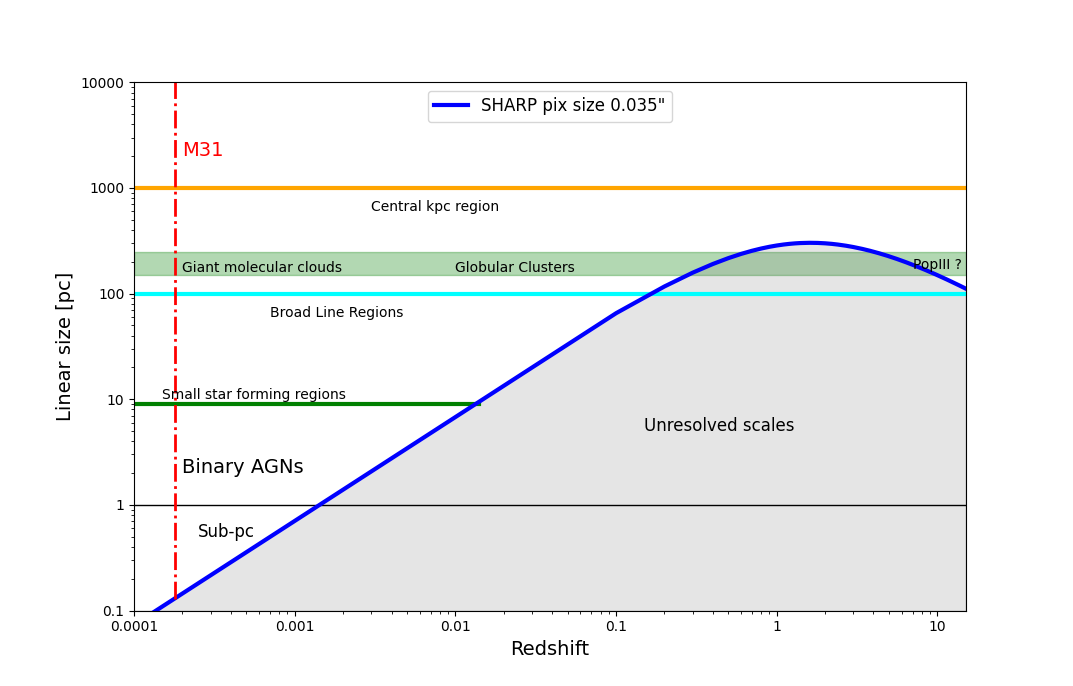}
    \caption{Left - Linear size [pc] subtended by a SHARP pixel (0.035$^{\prime\prime}$, blue curve) as a function of redshift compared to JWST NIRSpec (0.1"/pix, red curve). 
    The green stripe represents the typical size of giant molecular clouds (150-250 pc). Right - Same as Left panel but in logarithmic scale. The orange line marks 1 kpc scale, the green stripe represents the typical size of giant molecular gas clouds, 150-250 pc, globular clusters and high-redshift clumps observed by JWST. The green stripe also indicates an hypothetical PopIII system at $z\sim10$. Cyan and green lines mark the typical sizes of Broad Line Regions (BLR) and small nearby star forming regions respectively. 
The vertical dot-dashed red line marks the spatial resolution that can be
achieved at a distance similar to M31 galaxy.}
    \label{fig:linearscale}
\end{figure}

Typically, GMCs have sizes of $150$ to $250 \text{ pc}$ in nearby galaxies.
To sample these physical sizes across the entire cosmic timeline, an angular resolution of approximately $30-35$ mas is necessary. 
This is shown in Fig. \ref{fig:linearscale} (left panel) where the linear size 
subtended by an angle of $\sim$0.035 arcsec is shown as a function of redshift. 
This angular resolution ensures to resolve sub-galactic regions where massive star 
formation, enrichment and quenching processes take place, along the whole cosmic time.

% -------- SUGGESTED FIGURE PLACEHOLDER --------
% ----------------------------------------------
\subsection{Near-IR Coverage (0.95--2.45 $\mu$m)}
Tracking galaxy properties across cosmic time relies on analyzing specific atomic emission and absorption lines, e.g., Mgb and Fe tracing the metallicity, H$\alpha$ and [OII] diagnostics of star formation and AGN, Ca H\&K suited to estimate stellar kinematics, and the 4000\AA\ break mainly related to the mean age of the stellar population. 
Most of these key features are redshifted into the near-IR regime when observed in galaxies at $z>1-1.5$. 
Crucially, extending the coverage beyond $1.8 \mu$m is mandatory to search for 
Pop III stars via the HeII line at $z > 10$. 
The chosen upper limit of 2.45 $\mu$m is the optimal cut-off point for ground-based observations to minimize thermal background while maintaining high sky transmission.

\subsection{Sensitivity and Multiplexing}
Measuring stellar population gradients in high-redshift passive galaxies requires achieving a high Signal-to-Noise ratio (S/N $\sim$ 10) on the continuum at surface brightnesses down to $\mu_K \sim 22.0$ mag/arcsec$^2$ (see Saracco et al. 2026). 
For line emissions, detecting a hypothetical Pop III cluster of $10^6 M_\odot$ at $z\sim10$ requires reaching a line flux limit of $\sim 4 \times 10^{-20}$ erg/s/cm$^2$ 
(see Saracco et al. 2026). 
SHARP is required to reach these depths within a standard 10-hour integration. 
Furthermore, doing this efficiently over cluster fields or star-forming regions 
necessitates a high degree of multiplexing over a wide Field of View (FoV).

\section{Sharp architecture and optical design}
\label{sec:architecture}
\begin{figure}[htbp]
    \centering
    \includegraphics[width=4truecm]{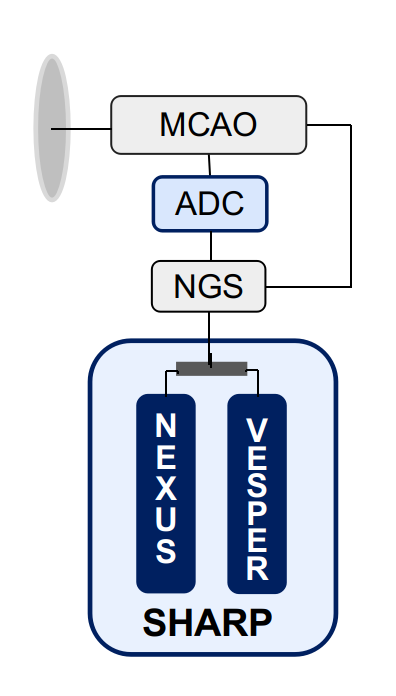}
    \hskip 2truecm
     \includegraphics[width=10truecm]{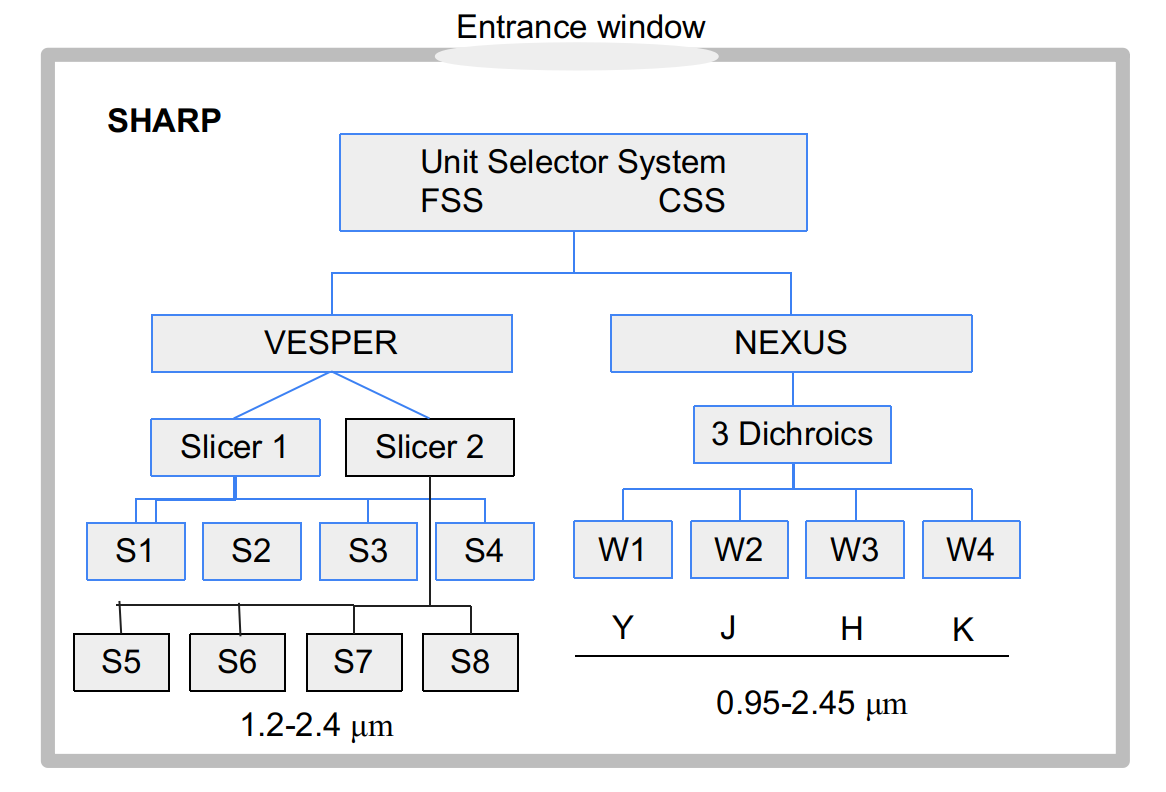}
    \caption{\label{fig:schema_sharp} Left -
   Schematic view of the components and of the optical path from the last mirror of the telescope (gray mirror on the left) to SHARP. 
   SHARP needs a MCAO unit, an NGS unit and an Atmospheric Dispersion Corrector (ADC).
   Note that the NGS unit is a component of the MCAO system.
  Right - SHARP architecture. SHARP is enclosed in a cryogenic tank. 
  Below the entrance window of SHARP there is the Unit Selector System (USS) that manages 
  the light between NEXUS and VESPER. NEXUS - Three dichroics split the beam into four wavelength ranges ($\approx$ Y, J, H and K) feeding 4 cameras, W1-W4. 
  The whole wavelength range 0.95-2.45 $\mu$m  is thus simultaneously covered.
VESPER - The image focused on the VESPER focal plane of one module is divided into 
four equal stripes (0.375" with each) by the slicer each one fed by a camera (S1-S4). 
In the current configuration VESPER hosts two modules to double the area probed. }
    \end{figure}

SHARP is enclosed in a cryogenic tank and operates downstream of the ELT's MCAO system (MORFEO) and an Atmospheric Dispersion Corrector (ADC; see Fig. \ref{fig:schema_sharp}). 
The light corrected by MORFEO is intercepted by the Unit Selector System (USS), 
which switches the beam between the two main scientific units: NEXUS and VESPER.
Below, we give an overview of the SHARP's functionalities. 
See \cite{saracco24} and \cite{mahmoodzadeh25} for a detailed description of the systems.

% -------- SUGGESTED FIGURE PLACEHOLDER --------

%\begin{figure}[htbp]
%    \centering
%    \includegraphics[width=6truecm]{SHARP_FOV.png}  
%    \caption{
%    The Field of View of NEXUS (blue square, $1.2^\prime \times 1.2^\prime$) and the area probed by the 12 Field Selectors of VESPER (red rectangle, $\sim 21^{\prime\prime} \times 40^{\prime\prime}$) superimposed on the AO-corrected area of the MCAO unit MORFEO (black circle, diameter $\sim 160^{\prime\prime}$).}
%    \label{fig:sharp_fov}
%\end{figure}
% ----------------------------------------------
\subsection{NEXUS: Multi-Object Spectrograph}
NEXUS operates over an AO-corrected FoV of approximately $1.2^\prime \times 1.2^\prime$ 
with a pixel scale of 35 mas/pixel. 
It uses a Configurable Slit System (CSS) capable of deploying up to 30 slits of 
minimum length $\sim 2.4^{\prime\prime}$. 

The light exiting the CSS encounters three dichroic lenses that split the beam into four distinct wavelength channels: 0.95--1.15 $\mu$m, 1.15--1.45 $\mu$m, 1.45--1.9 $\mu$m, and 1.9--2.45 $\mu$m. Each channel is fed into a dedicated camera equipped with high-performance grisms, yielding simultaneous full-band coverage (see Fig. \ref{fig:schema_sharp}). 
The system provides multiple spectral resolutions for extended sources ($R \approx 300, 2000, 6000$) depending on the science case (continuum detection vs. precision kinematic/abundance measurements), and achieves up to $R \sim 17000$ for point sources.

\subsubsection{The Inversion Prism}
A highly unique feature of the NEXUS CSS is the inclusion of an inversion prism (e.g., an Abbe-K\"onig prism) for each configurable slit. This optical element can rotate the square field of $\sim 2.4^{\prime\prime} \times 2.4^{\prime\prime}$ subtended by the slit by an arbitrary angle selected by the user. 

% -------- SUGGESTED FIGURE PLACEHOLDER --------
\begin{figure}[htbp]
    \centering
    \includegraphics[width=0.9\textwidth]{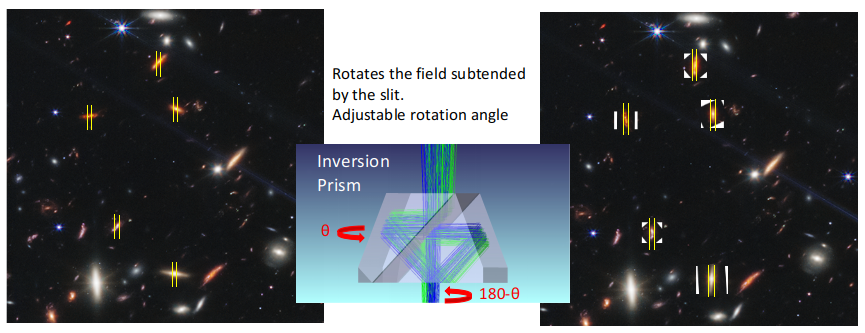}
    \caption{Optical concept of the inversion prism inside the NEXUS Configurable Slit System. Rotating the prism allows the user to align randomly oriented target galaxies along their major axes simultaneously within their respective slits. Note that in the center a Pechan prism is shown only for clarity while the current system design includes an Abbe-König prism.}
    \label{fig:prism}
\end{figure}
% ----------------------------------------------

By rotating the field, astronomers can align the major axis of any target along 
the slit, regardless of its original orientation on the sky. 
This capability is unprecedented in multi-object spectrographs.
In the example shown in Fig. \ref{fig:prism} it allows for the simultaneous 
derivation of rotation curves and dynamical masses for dozens of galaxies at once,
while such measurement could be carried out just for one galaxy at atime even by JWST. 
We observe that the prism implies a negligible transmission loss ($<$ 2\%) related 
to the two air-glass surfaces, one at the inlet and the other at the outlet thanks to
anti-reflective coatings.

\subsection{VESPER: Multi-Integral Field Unit}
VESPER is a modular image slicer designed to simultaneously cover the 1.2--2.4 $\mu$m wavelength range at a resolution of $R \sim 3000$. In its current configuration, VESPER comprises two modules, totaling 12 independent Field Selectors (FSs). 

\subsubsection{Field Selector System and Slicer}
Each FS has a FoV of $1.7^{\prime\prime} \times 1.5^{\prime\prime}$ and operates on a Cartesian grid, patrolling an area of approximately $21^{\prime\prime} \times 40^{\prime\prime}$. A system of prisms and movable collimators ensures a constant optical path length, allowing all 12 FSs to share the same focal plane seamlessly.

The light delivered by the FSS is focused onto a slicer array conceptually similar to the VLT/MUSE slicer. The image is divided into microscopic stripes, achieving a spatial sampling of 31 mas per spaxel. This high-density spatial sampling is specifically designed to maximize the S/N per spaxel given the encircled energy profile of the MORFEO MCAO correction. The output of the slicer is fed into eight $4\text{k} \times 4\text{k}$ detectors.

Table \ref{tab:properties} summarizes the main properties of NEXUS and VESPER.
\begin{table*}
\caption{- Properties of the Multi-Object Spectrograph NEXUS and the Multi-Integral Field Unit VESPER
of SHARP.} 
\label{tab:properties}
\begin{center}       
\begin{tabular}{|l|c|c|} %% this creates two columns
%% |l|l| to left justify each column entry
%% |c|c| to center each column entry
%% use of \rule[]{}{} below opens up each row
\hline
 {  SHARP property} &  {  NEXUS}&  {  VESPER}\\
\hline\hline
Spectral range (simultaneous) & 0.95-2.45 $\mu$m  & 1.2-2.4 $\mu$m \\
\hline\hline
FoV/Area probed & 1.2'$\times$1.2'  & 21"$\times$40"\\
\hline
Multiplexing & up to 30 slits (2.4" slit length)& 12 FSs (1.7"$\times$1.5" each)\\
\hline
Pixel scale & 35 mas  & 31 mas \\
\hline
Spectral resolution (ext. source) & 300, 2000, 6000 & 3000 \\
\hline 
Spectral resolution (point source) & $\sim$17000 & $\sim$4000 \\
\hline 
Number of cameras & 4 & 8 \\
\hline
Number of Detectors & 4 per camera& 1 per camera\\ 
\hline
Matrix & 2kx2k & 4kx4k \\
\hline
\end{tabular}
\end{center}
\end{table*}
\section{AN ILLUSTRATIVE SCIENCE CASE}
\label{sec:science_case}
To demonstrate the versatility of SHARP, we consider the high-redshift galaxy GLASS-180009 ($z \sim 2.66$). This system is massive ($4 \times 10^{10} M_\odot$), old ($\sim 1.7$ Gyr), and highly quiescent. Remarkably, observations show evidence of a massive neutral gas inflow ($\sim 19 M_\odot$/yr), yet no star formation is actively occurring
\citep{bevacqua26}. The galaxy resides within a larger overdensity.

% -------- SUGGESTED FIGURE PLACEHOLDER --------
\begin{figure}[htbp]
    \centering
    \includegraphics[width=0.6\textwidth]{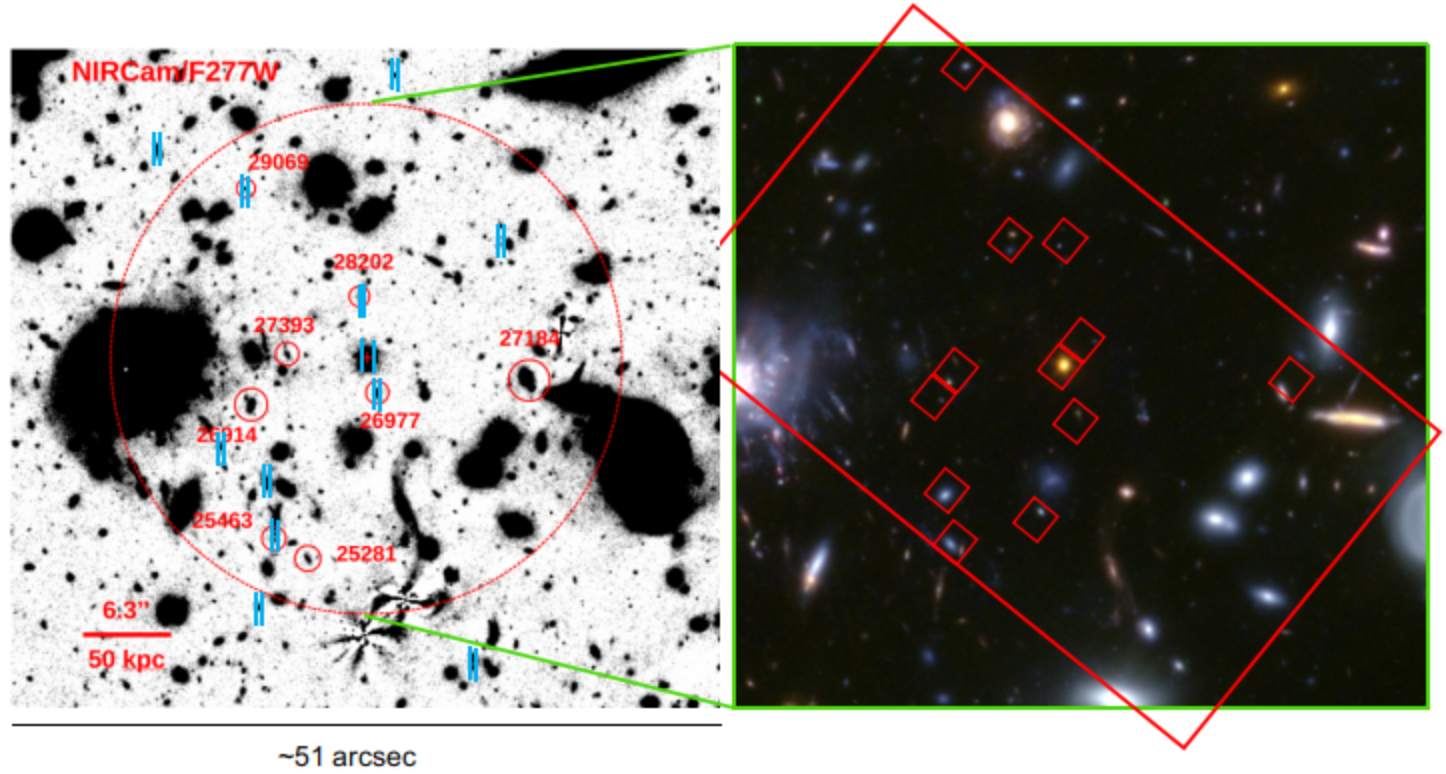}
    \includegraphics[width=0.3\textwidth]{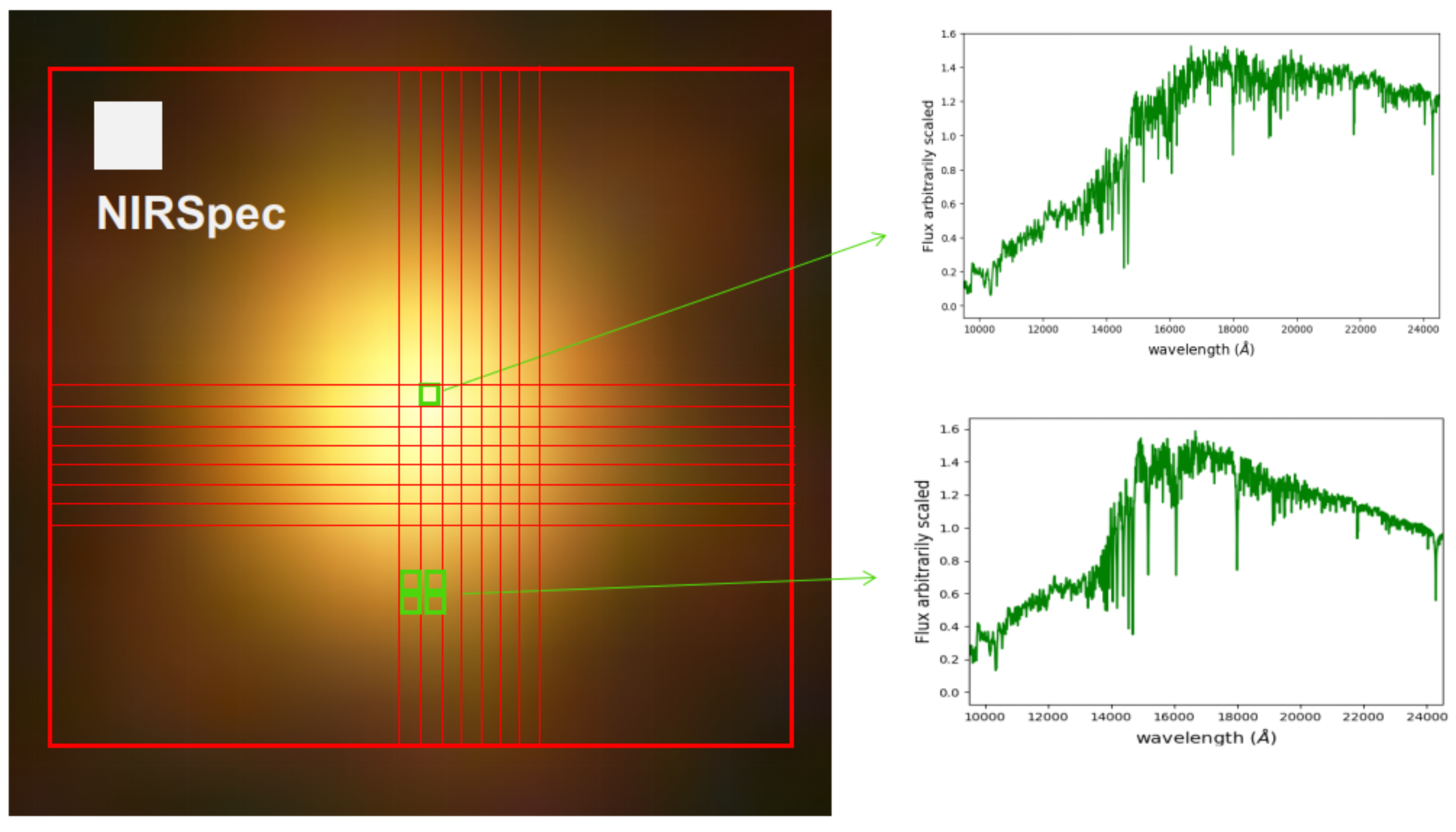}
    \caption{JWST NIRCam image of the field centered on GLASS-180009 ($z\sim2.66$). The left image illustrates how NEXUS slits (blue) can target the surrounding overdensity galaxies, while the VESPER Field Selectors (red squares, left) 
    perform high-resolution 3D spectroscopy on the central galaxy and its immediate circumgalactic environment simultaneously. The right panel shows the field of view of one FS (1.7"x1.5") centered on the central galaxy. The gray filled square represent the pixel size (0.1") of NIRSpec compared to the pixel size of VESPER (green squares, 0.031").}
    \label{fig:jades}
\end{figure}
% ----------------------------------------------

To unravel this mystery, one must probe the kinematics of the inflowing gas, 
map the stellar population gradients of the host galaxy, and characterize the surrounding environmental overdensity simultaneously. 
SHARP's architecture perfectly matches this challenge: 
NEXUS simultaneously targets the central and the surrounding galaxies within 
the $1.2^\prime \times 1.2^\prime$ field, utilizing the inversion prisms to 
align slits along their kinematic axes to confirm cluster membership, derive stellar
population properties (age, metallicity) and dynamical states.
VESPER probes the central galaxy at scales of $\sim 250$ pc to map the stellar gradients (age, metallicity) and dissect the gas inflow dynamics (Fig. \ref{fig:jades}, right).

Neither JWST, due to limited spatial resolution and fixed slit, nor the 
currently planned ELT suite, which lacks wide-field MCAO multiplexing capabilities, 
can execute this observation in a single, highly efficient configuration.

\section{The SHARP Science Book}
\label{subsec:science_book}
Beyond the overview on the scientific drivers given in Sec. \ref{sec:science} 
and the illustrative case discussed above (Sec. \ref{sec:science_case}), 
a broader collection of detailed reference science cases is being compiled 
in the SHARP Science Book, a Special issue of New Astronomy journal accessible at 
\url{https://www.sciencedirect.com/special-issue/10W3B6JJBXH}. 
These cases encompass a variety of astrophysical themes, ranging from 
AGN and black hole growth and co-evolution \citep[][Severgnini et al. 2026; Polletta et al. 2026]{vietri26}, to galaxy evolution and environmental effects across cosmic 
time \citep[e.g.,La Barbera et al. 2026;][]{gargiulo26,mancini26,polletta26a}, to very early Universe
\citep{bisogni26}. 
The reader is referred to the dedicated papers within this Special Issue 
of New Astronomy Journal for a comprehensive analysis of individual programs.
We note that the list above is incomplete since the compilation of this special issue 
is currently ongoing and some of the papers are still in the publication process.

\section{The SHARP Organization}
The SHARP project is currently in an important phase of growth and development. 
Unlike fully established collaborations, we are actively in the process of 
building a SHARP Consortium. 

{\it We warmly invite interested institutes, as well as individual researchers from the worldwide astronomical community, to join and participate in the project.} 

\begin{figure}[htbp]
    \centering
    \includegraphics[width=0.6\textwidth]{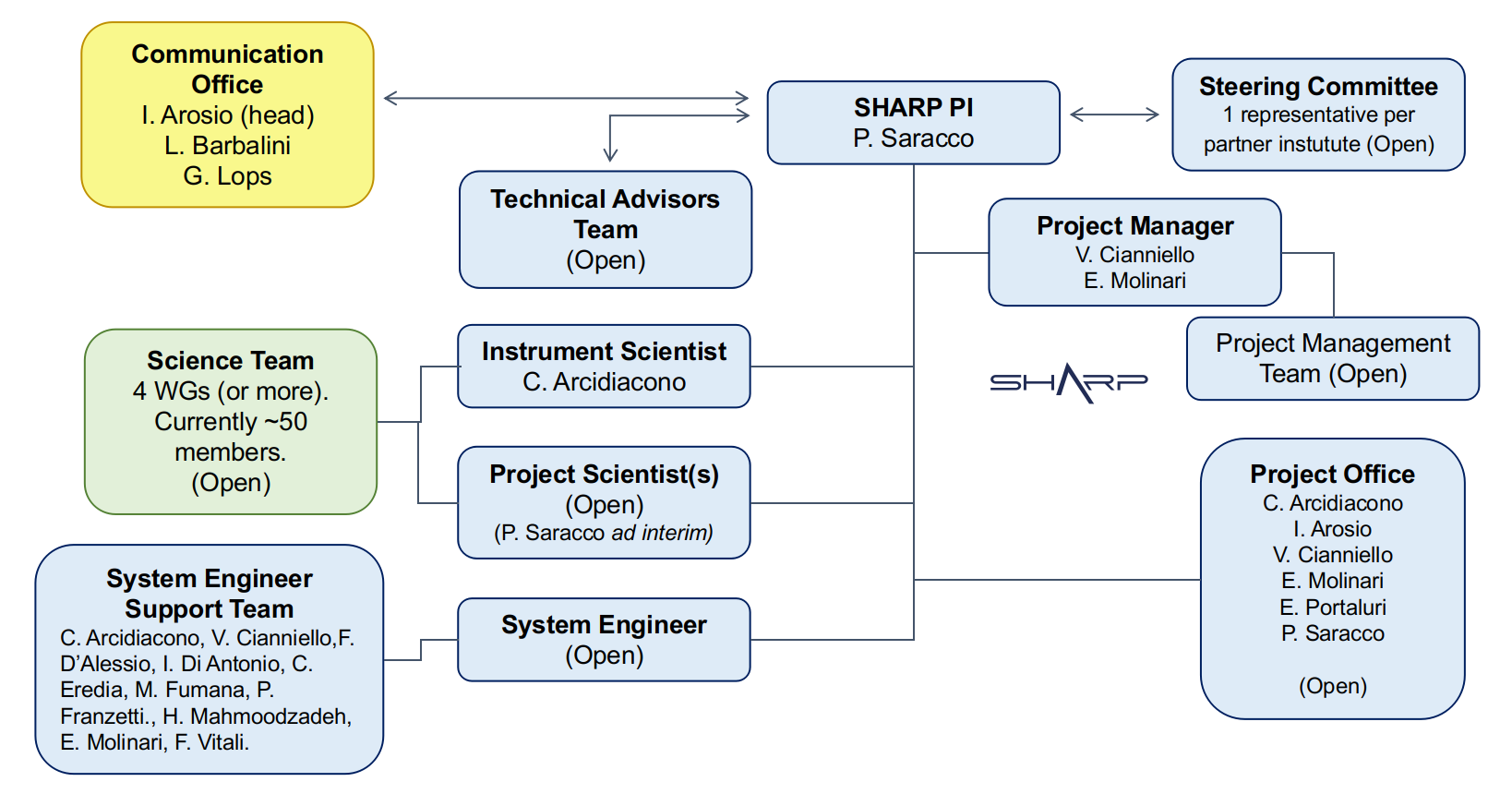}
    \caption{SHARP Organization Breakdown Structure (OBS). The SHARP project 
    is currently in a growth and development phase. 
We are interested in and actively working on establishing a SHARP consortium.
The label "Open" present in the OBS is an explicit signal of our 
    openness to integrate new partners also into key strategic 
    and technical roles.
    }
    \label{fig:obs}
\end{figure}

The current SHARP Organization Breakdown Structure (OBS) and the names of key 
persons involved in the project are shown in Fig. \ref{fig:obs}.
The large number of "TBD" (To Be Defined) present in our Organization Breakdown Structure (OBS) is not a lack of definition, but rather an explicit signal 
of our openness and willingness to integrate new partners also into key strategic 
and technical roles.
The project is represented by the Principal Investigator (PI), who has the 
ultimate responsibility for the project and serves as the primary formal contact point.
The Steering Committee (SC) will be the final decision-making body of the Consortium and the source of strategic guidance. 
It will be composed of one representative per partner institute. 
Currently, its composition is marked as TBD, pending the formalization of the 
joining partners.

Project Management assumes overall responsibility for the project. It will be supported by a Project Management Team, currently open to new contributions (to be defined). 

The Project Scientist (PS) and the Instrument Scientist lead the development 
of the science program and top-level requirements. 
The PS role is currently held ad interim by the PI, with the definitive position 
open for future assignment (TBD) as the consortium expands. 
All matters concerning the science cases are the responsibility of the PS and 
the Science Team (ST). The ST is currently composed of about 50 researchers 
and is organized into four Working Groups (WGs) each having a chair. 
The four WGs are: WG\#1 Early Universe and Cosmology (including  Early Universe, primordial galaxies, reionization, intergalactic/circumgalactic medium, absorption systems, gravitational lensing, dark matter, cosmological parameters); WG\#2 Galaxies
and AGNs (including  galaxies in their environment, structure, dynamics and kinematics,
interactions, unresolved stellar populations, assembly of galaxies, chemical evolution of galaxies; coevolution; physics of AGNs, supermassive BHs); WG\#3 Interstellar medium, Young Stellar Objects, Planetary Systems (including gas and dust, molecular clouds, HII regions, massive stars, (pre) MS stars, star formation, chemical processes in interstellar medium,  extrasolar planets); WG\#4 Resolved Stellar Population (including local Group, local volume, Milky way, Supernovae, pulsars, planetary nebulae,  AGB/postAGB stars, star clusters). 
We expect to expand the ST gathering members and expertise from all partner 
institutes. 
As the scientific team grows, we plan to increase the number of WGs, 
dedicating new groups to specific research topics that are currently grouped 
into broader thematic categories.

On the technical side, the System Engineer (SE) position is currently open to new partners (TBD). 
The future SE will be supported by an already established System Engineer Support Team (C. Arcidiacono, V. Cianniello, F. D’Alessio, I. Di Antonio, C. Eredia, M. Fumana, P. Franzetti, H. Mahmoodzadeh, E. Molinari, F. Vitali). 
The system engineering division works in close contact with the Instrument Scientist
(IS), who ensures that the adopted technical solutions perfectly match the foreseen
scientific needs of the instrument. 
Furthermore, a Technical Advisors Team (TBD) is planned to provide external or specialized guidance.

The Project Office ultimately and collegially encompasses all programmatic activities of the project. It is open to additional key members (+ TBD) representing future partner institutes.

Finally, recognizing the fundamental importance of communication, project dissemination, and internal coordination, SHARP has included a dedicated 
Communications Office from the outset. We invite you to visit the SHARP 
website and follow the project on LinkedIn.

\section{High-angular resolution near-IR spectroscopy in the 2030s-2040s }
Let us consider the landscape of near-IR high-angular resolution spectrographs and spectroscopy projected for the 2030s and 2040s.

Currently, NIRSpec \citep{jakobsen22} at JWST sets the standard for high-angular resolution Near-IR spectroscopy. 
The strength of NIRSpec on the JWST lies in operating outside the Earth's atmosphere,
which ensures a uniformly diffraction-limited field of view. 
As a result, its Micro-Shutter Assembly (MSA) can perform high-angular-resolution multiplexing for approximately 100 sources, complemented by a single Integral Field Unit (IFU).
NIRSpec angular resolution is limited by the pixel size to 0.1 arcsec, 
while the MSA system prevents the centering for all the targets.

While a new generation of ground-based telescopes, the ELT\citep{gilmozzi07}, the Giant Magellan Telescope \citep[GMT,][]{matt06,burgett24}, and the Thirty Meter Telescope (TMT), will become operational during the 2030s, the spectroscopy capabilities will remain largely confined by NIRSpec's current paradigm.
 
The ELT's AO-supported instruments include MICADO \citep{davies21,davies18}, with longslit capability along the parallactic angle and, likely, a single IFU; HARMONI \citep{niranjan24}, a single IFU, and ANDES \citep{marconi24}, a single longlist and IFU with extremely high spectral resolution (R$\sim$100000). 
Similarly, at the GMT is expected the Integral Field Spectrograph (IFS),
%\citep[IFS][]{sharp}, 
and at the TMT a comparable instrument, IRIS.
All these instruments will provide spectra for a single object at a time.
The exception is the MOSAIC \citep{pello24} multi-object spectrograph, 
whose fiber-fed system limits the angular resolution to 0.2" (the fiber size) 
and the wavelength range to 1.8 $\mu$m.
Furthermore, the versatility of systems like ANDES and IRIS is inherently constrained by the requirement for very bright nearby guide stars, which are essential for their Single-Conjugate Adaptive Optics (SCAO) systems.
Consequently, when the JWST mission concludes, likely between 2035-2040, a critical 
gap will emerge: there will be no spectrographs capable of efficiently pursuing 
and detailing the scientific pathways pioneered by JWST.
\\
\\
SHARP is conceived to maximize the scientific return of the ELT by fully exploiting the wide, MCAO-corrected field of view provided by MORFEO. By combining high angular resolution ($\sim$30 mas), continuous near-IR wavelength coverage (0.95--2.45 $\mu$m), and a unique dual architecture consisting of an adjustable-slit MOS (NEXUS) and a multi-IFU (VESPER), SHARP fills a critical gap in the currently planned ELT 
instrumentation suite. 
It is uniquely positioned to bridge the gap between the local and distant Universe, directly looking beyond the discoveries of JWST, allowing ELT to 
address the two scientific questions that will guide research in the coming decades:
\begin{itemize}
\item How matter assembled to form the first stars, galaxies, and the Universe we observe today;
\item Whether we are alone in the Universe, or if habitable worlds exist out there.
\end{itemize}

\acknowledgments % equivalent to \section*{ACKNOWLEDGMENTS}       
The SHARP team acknowledges support by Bando Ricerca Fondamentale INAF 2022, Techno-Grant "SHARP" - 1.05.12.02.01 and Bando Ricerca Fondamentale INAF 2024, Large-Grant "SHARP" - 1.05.24.01.01. We also refer the reader to the broader collection of science cases compiled in the SHARP Science Book.
\bibliography{sharp}   % bibliography data in report.bib
\bibliographystyle{spiejour}   % makes bibtex use spiejour.bst

\end{document}